\documentclass[
aps, prd, reprint,
nofootinbib,
balancelastpage,
amsmath,amssymb,
]{revtex4-2}

\usepackage{graphicx}
\usepackage{subcaption}
\usepackage{hyperref}
\hypersetup{
    colorlinks=true,
    linkcolor=black,
    citecolor=blue,
    urlcolor=black
}
 
\newcommand{\lcdm}{{$\Lambda$CDM}}
\newcommand{\fax}{f_{\mathrm{ax}}}
\newcommand{\hinvMpc}{h\,\mathrm{Mpc}^{-1}}
\newcommand{\maeV}[1]{m_a=10^{#1}\,\mathrm{eV}}

\begin{document}

\title{\textbf{Constraints on Ultra-Light Axions \\ from the DESI DR1 Full Shape, Planck and ACT} 
}% 
\author{Francesco Verdiani${}^{a,b,c,d}$}
\author{Emanuele Castorina${}^{e,f}$}
\author{Ennio Salvioni${}^{g,h}$}
\author{Emiliano Sefusatti${}^{b,c,d}$}
\author{Matteo Viel${}^{a,b,c,d}\;$}\email{fverdian@sissa.it,emanuele.castorina@unimi.it, esalvioni@ifae.es,emiliano.sefusatti@inaf.it,viel@sissa.it}
\affiliation{\vspace{0.2cm}
${}^{a}$SISSA - International School for Advanced Studies, Via Bonomea 265, 34136 Trieste,  Italy\\
${}^{b}$Istituto Nazionale di Astrofisica, Osservatorio Astronomico di Trieste, via Tiepolo 11, 34143 Trieste, Italy\\
${}^{c}$Institute for Fundamental Physics of the Universe, Via Beirut 2, 34151 Trieste, Italy\\
${}^{d}$Istituto Nazionale di Fisica Nucleare, Sezione di Trieste,  via  Valerio  2,  34127 Trieste,  Italy\\
${}^{e}$Dipartimento di Fisica ``Aldo Pontremoli'', Universit\`a degli Studi di Milano, Via Celoria 16, 20133 Milan, Italy\\
${}^{f}$INFN, Sezione di Milano, Via Celoria 16, 20133 Milan, Italy\\
${}^{g}$Departament de F\'isica, Universitat Aut\`onoma de Barcelona, 08193 Bellaterra, Barcelona, Spain\\
${}^{h}$IFAE and BIST, Campus UAB, 08193 Bellaterra, Barcelona, Spain
}

\begin{abstract}
We present updated bounds on ultra-light axions (ULAs) as a subcomponent of dark matter, derived from the full-shape analysis of the DESI Data Release 1 galaxy power spectra combined with Cosmic Microwave Background (CMB) data from ACT and Planck. We focus on the mass window $10^{-32}\,\mathrm{eV}\leq m_a \leq 10^{-24}\,\mathrm{eV}$, employing state-of-the-art analysis methods rooted in the Effective Field Theory of Large Scale Structure. For the smallest masses, our joint analysis with DESI improves over CMB-only constraints by more than a factor of 2, establishing the most stringent limits to date. For instance, for $m_a \sim 10^{-29}\,\mathrm{eV}$ ULAs are constrained to be a fraction as small as $0.3\%$ of the total matter energy density. The DESI Luminous Red Galaxy sample shows a mild preference for an ULA subcomponent with $m_a \approx 10^{-26}\,\mathrm{eV}$, mirroring previous hints from BOSS, but this preference vanishes upon combination with CMB data. Probing the largest masses, $m_a\gtrsim10^{-25}\,\mathrm{eV}$, in future studies will benefit from extending the data analysis to smaller scales, both for galaxy clustering and CMB lensing, but will also require concurrent improvements in the theoretical modeling.
\end{abstract}

\maketitle

\section{Introduction}
Dark matter (DM) is a pillar of the standard cosmological model, but its fundamental nature has so far remained elusive. As all the present experimental evidence for DM stems from its gravitational interactions, the dark sector may be secluded from the Standard Model, possibly leaving astrophysics and cosmology as our only windows into its properties. These fields offer complementary perspectives: while astrophysics targets localized systems, yielding tests of the small-scale properties of the dark sector, cosmology provides robust and statistically grounded probes on the largest scales.

The anisotropies of the Cosmic Microwave Background (CMB) are a prime example of this robustness. As both the observational measurement and theoretical modeling are firmly established, they provide exquisite precision tests of the nature of DM, finding so far no hints of deviations from a purely cold and collisionless behavior. However, transformative progress on both the observational and theoretical sides is bringing galaxy clustering on the same ground, and in a very competitive position with the CMB. With the advent of Stage IV surveys such as the Dark Energy Spectroscopic Instrument (DESI), Euclid, and the Vera C. Rubin Observatory, the precision and volume of Large Scale Structure (LSS) datasets are reaching unprecedented levels. At the same time, the theoretical development of the Effective Field Theory of Large Scale Structure (EFTofLSS) \cite{Baumann:2010tm,Carrasco:2012cv} provides an analytical framework that allows one to obtain from first principles, within or beyond the standard cosmological model, a robust parameter inference, without phenomenological or poorly-controlled theoretical assumptions.

Ultra-light axions (ULAs) are compelling DM candidates for which the prime observable of spectroscopic galaxy surveys, namely the full shape of the galaxy power spectrum, is expected to add genuine new sensitivity. Originally proposed as a solution to the strong CP problem \cite{Peccei:1977hh,Weinberg:1977ma,Wilczek:1977pj}, axions as dark sector components are also motivated from string theory~\cite{Svrcek:2006yi,Arvanitaki:2009fg}. For what concerns the cosmological evolution, an important feature of axions is that once the Hubble parameter becomes comparable to the axion mass, $H(z) \sim m_a$, the field begins to coherently oscillate~\cite{Preskill:1982cy,Abbott:1982af,Dine:1982ah} and its background energy density redshifts as nonrelativistic matter from that time onwards. As a result, axions could form a component of DM today provided that $m_a \gtrsim 10^{-32}\,\mathrm{eV}$ (even lighter axions would behave as dark energy today).

\begin{figure*}[t]
    \centering
    \includegraphics[width =\textwidth]{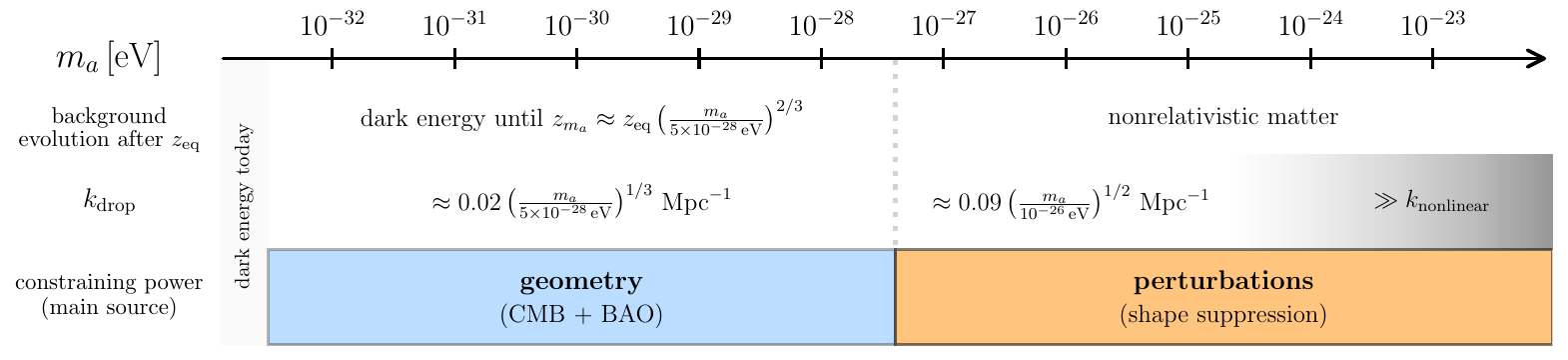}
    \caption{Schematic representation of the cosmological features of ULAs, in the range of masses for which the suppression scale $k_\mathrm{drop}$, as defined in Eq.~\eqref{eq:k_drop}, is on cosmological scales (middle row). Which observables carry the strongest observational imprints (bottom row) mainly depends on the background evolution after the redshift of matter-radiation equality, $z_{\rm eq}$ (top row). This work focuses on $m_a\leq 10^{-24}\,\mathrm{eV}$, where a perturbative modeling of the observables is possible.}
    \label{fig:ULAs_schematic}
\end{figure*}

What makes the ultra-light part of the axion DM mass range especially interesting, is that ULAs, while matching the background evolution of standard cold DM (CDM), exhibit deviations from a cold species at the level of cosmological perturbations. Indeed, due to their macroscopic de Broglie wavelength, axions develop a characteristic, time-dependent scale below which the growth of perturbations is suppressed~\cite{Hu:2000ke,Amendola:2005ad,Marsh:2010wq,Hlozek:2014lca}. This physical scale decreases for heavier axions and along the expansion of the Universe, leading to an upper bound on $m_a$ for ULAs to leave discernible imprints on cosmological observables. Broadly speaking, the suppression is imprinted on astrophysical/cosmological scales if $m_a\lesssim10^{-18}\,\mathrm{eV}$, and in particular it can be probed with galaxy clustering and CMB (lensing) data if $m_a\lesssim10^{-24}\,\mathrm{eV}$. 

Hence, there exists a mass window of about $8$ orders of magnitude, $10^{-32}\,\mathrm{eV}\leq m_a \leq 10^{-24}\,\mathrm{eV}$, where a DM abundance made of ULAs can be probed with a joint analysis of CMB and spectroscopic galaxy clustering. The overall consistency between data and the predictions of pure CDM implies that ULAs in this mass window cannot constitute the totality of DM. Nevertheless, they may form a subcomponent of DM, and that is the possibility we focus on here, as sketched in Fig.~\ref{fig:ULAs_schematic}. Recent data from DESI~\cite{DESI:2023ytc,DESI:2024aax,DESI:2025fxa} and the Atacama Cosmology Telescope (ACT)~\cite{ACT:2023dou,ACT:2023kun,AtacamaCosmologyTelescope:2025blo,AtacamaCosmologyTelescope:2025nti} stand out for their potential to probe this scenario.

The latest Data Release 1 (DR1) from DESI~\cite{DESI:2024aax,DESI:2025fxa} provides a timely opportunity to test the presence of an ULA DM subcomponent. Comprising spectroscopic measurements of luminous red galaxies (LRG), emission line galaxies (ELG) and quasars (QSO), as well as a sample of bright galaxies (BGS), DR1 maps nearly 6 million LSS tracers spanning an effective cosmic volume of approximately $18\,\mathrm{Gpc}^3$. This is an improvement by a factor of $\sim 3$ compared to SDSS DR12~\cite{eBOSS:2020yzd} data, which were used to derive previous constraints on ULAs from the full shape of the galaxy power spectrum~\cite{Verdiani:2025jcf}, see also~\cite{Lague:2021frh,Rogers:2023ezo} for earlier results. The official DESI analysis~\cite{DESI:2024hhd,DESI:2024mwx} and later reanalyses~\cite{Chudaykin:2025aux,Chudaykin:2025lww,Chudaykin:2025vdh} of DR1 full-shape data within the EFTofLSS framework have demonstrated its remarkable constraining power, but have so far been restricted to the baseline {\lcdm} model and a few of its standard extensions. Given the pristine quality of the data, there is strong motivation to push these boundaries and exploit this dataset as a laboratory to probe fundamental physics.

In this paper, we therefore present the first application of DESI DR1 full-shape data to constrain non-cold components of the dark sector, adopting ULAs as a benchmark scenario. Rather than relying on standard pipelines hard-coded for {\lcdm}, we perform an independent reanalysis of the DESI dataset. For this purpose, we employ a self-consistent EFTofLSS framework explicitly tailored to capture the signatures of mixed (i.e., cold + non-cold) DM models, based on our previous work~\cite{Verdiani:2025jcf}. 
Modeling the nonlinear dynamics, redshift-space distortions, and galaxy bias from first principles ensures that the distinctive beyond the Standard Model dynamics of ULAs is consistently isolated from, and not mistaken for, poorly controlled astrophysical or nonlinear effects. Furthermore, by combining this LSS pipeline with ACT DR6~\cite{ACT:2023dou,ACT:2023kun,AtacamaCosmologyTelescope:2025blo,AtacamaCosmologyTelescope:2025nti} and Planck~\cite{Planck:2018vyg,Planck:2019nip,Carron:2022eyg} data, we are able to break background parameter degeneracies and isolate the clustering signal.

The rest of this paper is organized as follows. In Sect.~\ref{sect:ULA-theory} the essential features of ULAs are reviewed, focusing on the linear cosmology evolution. Section~\ref{sect:data-and-method} then presents the datasets employed in the present study, and the methods with which they were analyzed. Our main results, including updated upper limits on the abundance of an ULA DM component as a function of the ULA mass, are presented in Sect.~\ref{sect:results}. We focus first on CMB data alone, then on the DESI DR1 full shape, and finally on their combination. Section~\ref{sect:conclusions} highlights the outcomes of this work and outlines future directions. Appendices~\ref{sect:lensing-ULAs-1loop} and~\ref{sect:DESI-lcdm} provide further technical details on our analysis.

\section{Linear cosmology of ultra-light axions}\label{sect:ULA-theory}
\begin{figure*}[t]
    \centering
    \begin{subfigure}{.49\textwidth}
     \includegraphics[width=3.4in]{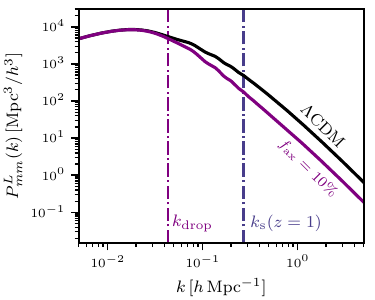}
    \end{subfigure}
    \hfill
    \begin{subfigure}{.49\textwidth}
        \includegraphics[width=3.4in]{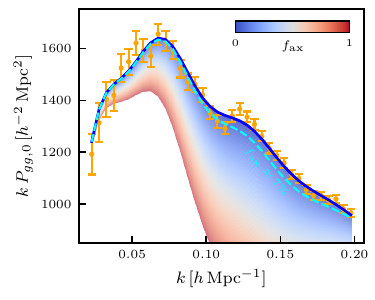}
    \end{subfigure}
    \caption{\textit{Left}: Linear matter power spectrum at $z = 1$ in an ULA cosmology with $\maeV{-27}$ and $f_{\rm ax} = 10\%$, compared to $\Lambda\mathrm{CDM}$. The two relevant scales $k_{\rm drop}$ and $k_s(z)$ discussed in Sect.~\ref{sect:ULA-theory} are marked by vertical dot-dashed lines. \textit{Right}: Monopole of the galaxy power spectrum for the \texttt{LRG1} bin of DESI DR1, chosen for illustration. Points with errorbars correspond to the data, while lines show theoretical predictions for an ULA cosmology with $\maeV{-26}$. The blue line is the best fit, which is compatible with pure CDM, having negligible $\fax$. The predictions for cosmologies with increasing $f_\mathrm{ax}$ are shown by the color gradient, whereas the cyan dashed line corresponds to the largest value allowed by data (DESI-only analysis) at $95\%$ CL, $f_{\rm ax}\approx 7\%$. When $\Omega_a h^2$ is increased, $\Omega_{\rm cdm} h^2$ is reduced keeping their sum fixed; for illustrative purposes, all other parameters are kept fixed.}\label{fig:ULA-linear-power}
\end{figure*}
ULAs are described by a scalar field $\phi$, minimally coupled to gravity but otherwise decoupled from all the other cosmic species, with a potential that is well approximated by a pure mass term, $V(\phi) = m_a^2 \phi^2 / 2\,$.\footnote{Following the implementation in the Boltzmann solver \texttt{axiCLASS}~\cite{Poulin:2018dzj,Smith:2019ihp}, the ULA potential is modeled as $V(\phi) = m_a^2 f_a^2 [1 - \cos(\phi/f_a)]$, where $f_a$ is set to $0.4\hspace{0.2mm} M_{\rm Pl}$. We have verified that this produces negligible differences with respect to a pure mass term in the parameter space we consider.} The evolution of the background value of the field, $\overline{\phi}$, is governed by the Klein-Gordon equation
\begin{equation}
\overline{\phi}^{\,\prime\prime} + 2 \mathcal{H} \overline{\phi}^{\,\prime} + a^2 m_a^2 \overline{\phi} = 0\,,
\end{equation}
where primes denote derivatives with respect to conformal time and $\mathcal{H} \equiv a H$, with $a$ denoting the scale factor. The initial conditions are $\phi(a_i)$, which is traded in practice for the present-day fraction of energy density in ULAs, $\Omega_a h^2$, and $\overline{\phi}^{\,\prime}(a_i) = 0$. At first, the field remains stuck due to Hubble friction and acts as a dark energy component, $\bar{\rho}_a \propto  a^0$. Once the expansion of the Universe has slowed down to a sufficient degree, namely around $a = a_{m_a}$ with $3 H(a_{m_a}) = m_a$, the field begins to oscillate rapidly around the minimum of its quadratic potential. Upon averaging over the coherent oscillations, the energy density in ULAs redshifts as nonrelativistic matter, $\bar{\rho}_a \propto a^{-3}$. 

In an effective fluid description for the linear cosmological perturbations, in the matter-dominated era and on sub-horizon scales ($k \gg \mathcal{H}$), the Euler equation for the axion component develops a sound speed term with $c_s^2 = k^2 / (4m_a^2 a^2)$~\cite{Hu:2000ke,Hlozek:2014lca,Urena-Lopez:2015gur,Cookmeyer:2019rna,Passaglia:2022bcr}. Therefore, ULA perturbations cannot grow below the characteristic (Jeans) scale
\begin{equation}\label{eq:Jeans_ULAs}
k_s(z) \approx  \frac{0.069}{\mathrm{Mpc}} \bigg( \hspace{-0.4mm}\frac{m_a}{10^{-28}\,\mathrm{eV}}\hspace{-0.4mm}\bigg)^{\hspace{-1mm}1/2} \hspace{-0.75mm} \left( \frac{\Omega_m^0 h^2}{0.14} \right)^{\hspace{-1mm}1/4}\hspace{-0.75mm} (1+z)^{-1/4}.
\end{equation}
In the mixed DM scenario we consider, ULAs form a DM subcomponent gravitationally coupled to the CDM+baryon fluid (denoted in this work by $c$; cold dark matter alone is denoted by $\mathrm{cdm}$), with fractional matter energy density 
\begin{equation}
f_{\rm ax} \equiv \frac{\Omega_a}{\Omega_c + \Omega_a}\,.
\end{equation}
The linear matter power spectrum, $P_{mm}^L$ (where the matter field $m$ includes both CDM+baryons and ULAs), shows a step-like suppression as a consequence of the Jeans scale, as shown in the left panel of Fig.~\ref{fig:ULA-linear-power}. The fraction $f_{\rm ax}$ determines the depth of the step, whereas the mass $m_a$ sets the location: the smallest wavenumber affected by the suppression is $k_\mathrm{drop} = k_s (\mathrm{min}\hspace{0.2mm}\{ z_{m_a}, z_{\rm eq}\})$, or
\begin{equation} \label{eq:k_drop}
k_{\rm drop}\hspace{-0.7mm} \approx \hspace{-0.85mm} \begin{cases} 
\hspace{-0.6mm}\frac{0.091}{\mathrm{Mpc}}\hspace{-0.2mm} \Big( \hspace{-0.4mm}\frac{m_a/\mathrm{eV}}{10^{-26}}\hspace{-0.4mm}\Big)^{\hspace{-0.75mm}1/2}, \;\qquad\quad\;\,\hspace{0.2mm} \frac{m_a}{\mathrm{eV}} \hspace{-0.45mm}>\hspace{-0.5mm} 5 \hspace{-0.5mm}\times \hspace{-0.6mm} 10^{-28} \\
\hspace{-0.6mm}\frac{0.012}{\mathrm{Mpc}}\hspace{-0.2mm} \Big( \hspace{-0.4mm}\frac{m_a/\mathrm{eV}}{10^{-28}}\hspace{-0.4mm}\Big)^{\hspace{-0.75mm}1/3} \hspace{-1.25mm} \left( \hspace{-0.5mm}\frac{\Omega_m^0 h^2}{0.14}\hspace{-0.5mm} \right)^{\hspace{-0.75mm}1/3}\hspace{-1.5mm},\frac{m_a}{\mathrm{eV}} \hspace{-0.45mm}<\hspace{-0.5mm} 5 \hspace{-0.5mm}\times \hspace{-0.6mm} 10^{-28}
\end{cases}
\end{equation}
depending on whether the ULAs start to oscillate during radiation or matter domination, respectively. The step in the matter power spectrum flattens out above $k_s(z)$, where $z$ is the redshift of observation. For $k\gg k_s(z)$, the ULA perturbations are completely negligible, while the linear growth rate of the cold fluid is suppressed by $1-3 f_\mathrm{ax}/5$ relative to the {\lcdm} expectation. Figure~\ref{fig:ULAs_schematic} sketches how the constraining power of different cosmological observables depends on the value of $m_a$. A broader class of such mixed DM scenarios have been recently explored in~\cite{Amin:2025nxm}.

\section{Data and Methods}\label{sect:data-and-method}

\subsection{Datasets}
\paragraph{CMB} We employ the state-of-the-art combination of ACT Data Release 6 (DR6) \cite{ACT:2023dou,ACT:2023kun,AtacamaCosmologyTelescope:2025blo,AtacamaCosmologyTelescope:2025nti} and Planck \cite{Planck:2018vyg,Planck:2019nip} legacy CMB datasets, supplemented by reconstructed lensing measurements. 
The optimal primary CMB combination that we consider, dubbed \textsc{P-ACT} in \cite{AtacamaCosmologyTelescope:2025blo}, merges the high-resolution temperature and polarization measurements from ACT with the large-scale full-sky observations from Planck. To avoid overlap, the Planck high-$\ell$ likelihood is explicitly truncated  at $\ell_\mathrm{max}=1000\,(600)$ for TT~(TE, EE) \citep{AtacamaCosmologyTelescope:2025blo}. 

Analogously, we make use of the CMB lensing combination of Planck PR4 + ACT DR6 in the same setup of the official ACT analysis \cite{ACT:2023kun,ACT:2023dou,Carron:2022eyg}. While Planck lensing covers a larger sky area, ACT probes smaller scales with better precision; we consider the \texttt{baseline} range of multipoles $40<L<763$, while fitting the \texttt{extended} spectrum (up to $L = 1300$)~\cite{ACT:2023ubw} is left to future work, for reasons outlined in what follows.

\vspace{1mm}
\paragraph{DESI DR1 full-shape}
Starting from the DESI DR1 catalogs~\cite{DESI:2024aax} we measured the power spectra with \texttt{pypower}\footnote{https://github.com/cosmodesi/pypower} \cite{Hand:2017irw}, which implements the $\theta$-cut, removing all pair counts with angular separation $\theta < 0.05^{\rm o}$~\cite{Pinon:2024wzd} to mitigate fiber assignment incompleteness. The numerical covariances have been measured from the approximate \texttt{EZmocks} catalogs \cite{DESI:2024jxi}. The mixing matrices relevant for the convolution of the theory model with the survey mask, are then obtained following standard methods \cite{DESI:2024jxi, Beutler:2018vpe} with the \texttt{convo}\footnote{https://jacoposalvalaggio.gitlab.io/convo} code.

\begin{figure*}[t]
    \centering
    \begin{subfigure}{.49\textwidth}
        \includegraphics[width=3.4in]{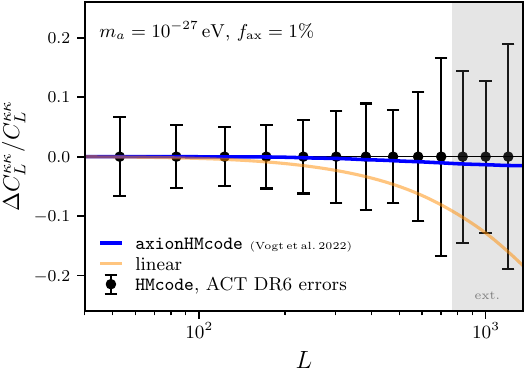}
    \end{subfigure}
    \hfill
    \begin{subfigure}{.49\textwidth}
     \includegraphics[width=3.4in]{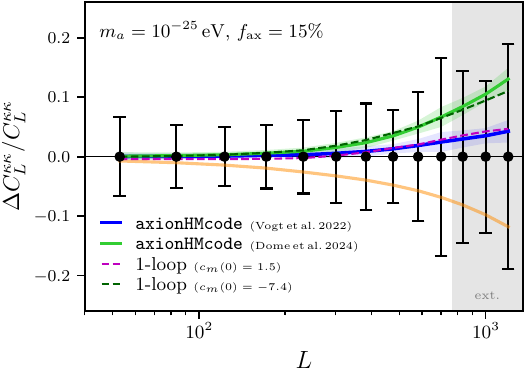}
    \end{subfigure}    
    \caption{Fractional difference between $C_L^{\kappa\kappa}$ computed with different prescriptions for the nonlinear corrections to $P_{mm}$, and the result obtained using the standard \texttt{HMcode}~\cite{Mead:2020vgs}. The ACT DR6 errorbars are also shown for reference.  Two values of the ULA mass are considered, while $\fax$ is set close to the upper limit allowed by the data in each case (see Table \ref{tab:result-uppers}). The extended range $763<L<1300$, not included in the present study, is shaded in gray. {\it Left:} $m_a = 10^{-27}\,\mathrm{eV}$. The solid blue line employs~\texttt{axionHMcode} in the version of~\cite{Vogt:2022bwy} for the nonlinear corrections. We do not show the version of~\cite{Dome:2024hzq}, because it was calibrated to simulations at the much larger mass $\maeV{-24.5}$. The linear result is reported in orange. {\it Right:} $m_a = 10^{-25}\,\mathrm{eV}$. Solid blue and green lines employ the two different versions of~\texttt{axionHMcode}~\cite{Vogt:2022bwy,Dome:2024hzq} for the nonlinear corrections. Dashed lines correspond to our EFTofLSS predictions, with the counterterm $c_m(0)$ adjusted to reproduce the \texttt{axionHMcode} results.}
    \label{fig:axionHMcode_lensing}
\end{figure*}

For each tracer, we make use of the $P_{gg,\hspace{0.2mm}\ell}$ multipoles with $\ell=0$ and $2$, discarding the hexadecapole due to the poor signal-to-noise ratio. We adopt a fixed scale cut of $k_\mathrm{max}=0.2\,\hinvMpc$ as in previous analyses~\cite{DESI:2024jxi,DESI:2024hhd,Chudaykin:2025aux}.
Compared to the official analysis, we do not apply a further rotation to the data vector, as it has been observed that the quite non-diagonal structure of the window matrix does not affect inference~\cite{DESI:2024jxi,Pinon:2024wzd}, performing the convolution up to $k_\mathrm{conv}=0.35\,\hinvMpc$ as discussed in \cite{Pinon:2024wzd}. We anyway tested the stability of our main results for the joint analysis against further increases of $k_\mathrm{conv}$, confirming the lack of appreciable differences in the cosmological parameter posteriors.
Further details are reported in Appendix~\ref{sect:DESI-lcdm}.

\subsection{CMB theoretical predictions}\label{sect:cmb-theory-predictions}
Linear power spectra are computed using the Boltzmann solver \texttt{axiCLASS}~\cite{Poulin:2018dzj,Smith:2019ihp}, which implements the ULA dynamics. However, CMB lensing is sensitive to nonlinearities in $P_{mm}$. The previous ULA analyses based on Planck+BOSS data~\cite{Rogers:2023ezo, Verdiani:2025jcf} (see also~\cite{Hlozek:2017zzf,Lague:2021frh}) computed the lensing potential power spectrum using {\it linear} predictions, which was shown not to bias the results for Planck errorbars. The ACT DR6 four-point function measurements from which the lensing potential is extracted, however, probe smaller~(hence, more nonlinear) scales with better precision, therefore this aspect needs to be revisited~\cite{Hlozek:2016lzm,Gaughan:2026xrv,Lague:2026sbd}.

The left panel of Fig.~\ref{fig:axionHMcode_lensing} compares two different predictions for the lensing convergence power spectrum, $C_L^{\kappa\kappa}$, both of which include nonlinear corrections through halo model-based fitting functions: the standard \lcdm-based \texttt{HMcode}~\cite{Mead:2020vgs} (to which we give the total $P_{mm}^L$, comprising the ULA component, as input) and \texttt{axionHMcode}, which was developed in~\cite{Vogt:2022bwy} specifically to include ULA nonlinearities. The ACT DR6 lensing errorbars are also displayed, for reference.
For sufficiently light ULAs, $m_a \lesssim 10^{-27}\,\mathrm{eV}$, the fraction $f_{\rm ax}$ is constrained to be subpercent and, as a consequence, genuine ULA nonlinearities are negligible for ACT scales and errorbars. On the other hand, we see that a linear prediction is not sufficient, as it entails a systematic error of the order of the statistical one. 

Larger ULA masses, for which the allowed fraction is larger, pose a theoretical challenge in the description of genuine ULA nonlinearities.
This is shown in the right panel of Fig.~\ref{fig:axionHMcode_lensing}, taking $\maeV{-25}$, for which $f_{\rm ax} = 15\%$ is allowed by the combination of CMB (excluding lensing) and DESI data. We see that, as already noted in~\cite{Gaughan:2026xrv}, the two \texttt{axionHMcode} halo-model implementations available for this heavier ULA regime~\cite{Vogt:2022bwy,Dome:2024hzq} yield different predictions. Furthermore,~\cite{Gaughan:2026xrv} has shown that cosmological inference of $\Omega_a h^2$ is highly sensitive to the choice of the halo-model-based nonlinear prescription, already with the \texttt{baseline} scale window $L<763$. This can even produce an artificial preference for ULAs with $m_a\approx 10^{-24}\,\mathrm{eV}$~\cite{Gaughan:2026xrv}. Hence, nonlinear modeling is a theoretical systematic uncertainty that, at present, challenges a robust exploration of the $m_a\gtrsim 10^{-25}\,\mathrm{eV}$ regime.

In this work, we therefore adopt a perturbative modeling of CMB lensing, focusing on multipoles with $L<763$, for which the bulk of the CMB lensing amplitude lies on quasi-linear scales. The 1-loop prediction for the matter power spectrum in the framework of the EFTofLSS is
\begin{equation}\label{eq:Pmm-1l-lensing}
\begin{split}
P_{mm}(k,z) =&\;P^{L}_{mm}(k,z) \\ 
+&\;P^\mathrm{1\text{-}loop}_{mm}(k,z) -2 c_m(z)\hspace{0.3mm}k^2P^{L}_{mm}(k,z)\,,
\end{split}
\end{equation}
where $P^\mathrm{1\text{-}loop}_{mm}(k,z)$ is the next-to-leading-order correction, while the last term is proportional to the EFT coefficient (counterterm) $c_m$ that absorbs any small-scale dependence, beyond the perturbative regime, of the loops. Notice that standard CDM and baryons are treated as indistinguishable on large scales. The challenge is that the broad redshift span of the CMB lensing kernel requires knowledge of the time dependence of $c_m$. To address this, several approaches have been put forward \cite{ Foreman:2015uva,Philcox:2020rpe,Braganca:2020nhv,Chen:2024vuf,Ivanov:2026dvl} that still allow one to use the bulk of information on large scales from perturbation theory. Here, as a proof of principle, we show that for ACT errorbars the 1-loop model above is already sufficient to capture these deviations from linear theory. Following~\cite{Foreman:2015uva}, we assume 
\begin{equation}\label{eq:cs2-timedep}
    c_m(z) = c_m(0) D(z)^p\,,
\end{equation}
where $D(z)$ is the linear growth factor, normalized to unity at $z = 0$. We take $p=3$ as motivated by comparison with N-body simulations \cite{Foreman:2015uva}. In the right panel of Fig.~\ref{fig:axionHMcode_lensing} we observe that by adjusting the value of $c_m(0)$ (red and dark green dashed lines) we are able to reproduce both halo-model prescriptions that include ULA nonlinearities, shown by the blue and light green solid lines. In this respect, our analysis is more conservative than the recent one in~\cite{Lague:2026sbd}, which employed the halo model, as we marginalize over one extra free parameter, namely the value of $c_m(0)$. Appendix~\ref{sect:lensing-ULAs-1loop} reports additional tests of our theoretical modeling, showing in particular that, on the scales considered here, the prediction is quite insensitive to the value of $p$ in Eq.~\eqref{eq:cs2-timedep}. In summary, for $m_a\geq 10^{-26}\,\mathrm{eV}$ we adopt the conservative EFTofLSS modeling, while for lighter ULAs ($m_a\leq 10^{-27}\,\mathrm{eV}$) the standard~\texttt{HMcode} is sufficient, due to the smallness of the allowed $f_{\rm ax}$.

Achieving a computationally efficient, yet accurate, evaluation of $P^\mathrm{1\text{-}loop}_{mm}(k,z)$ in Eq.~\eqref{eq:Pmm-1l-lensing} is a separate challenge, as the presence of the ULA Jeans scale increases the complexity of cosmological perturbation theory beyond linear order. For this purpose we apply the results of~\cite{Verdiani:2025jcf}, where we presented the extension of the EFTofLSS to general mixed DM scenarios, including cosmologies with a DM subcomponent made of ULAs. The framework of~\cite{Verdiani:2025jcf} is based on a perturbative expansion for small $f_{\rm ax}$, which is justified in the mass range where DESI data have sensitivity, $m_a\lesssim 10^{-25}\,\mathrm{eV}$. More details of this modeling are presented in Sect.~\ref{sect:eftoflss}, since they are shared by the predictions for galaxy clustering.

\subsection{Galaxy clustering model:\\ EFTofLSS with two fluids}
\label{sect:eftoflss}

The EFTofLSS provides the state-of-the-art modeling of the galaxy power spectrum. However, its original formulation~\cite{Baumann:2010tm,Carrasco:2012cv} is tailored to a pure-CDM, scale-free cosmology, and therefore unsuitable to perform inference in mixed DM scenarios, where one component has a characteristic Jeans scale.
For this purpose, in~\cite{Verdiani:2025jcf} we developed a two-fluid extension of the EFTofLSS. The main theoretical challenge stems from the fact that the standard Eulerian Perturbation Theory, upon which the EFTofLSS is based, relies on the factorization of the time-dependent growth factors to compute loop integrals. This factorization applies only to scale-free cosmologies. Extensions to scenarios with scale-dependent growth have been discussed in relation to massive neutrinos~\cite{Garny:2020ilv,Aviles:2021que,Noriega:2022nhf}; however, that can also be approximated as a scale-free problem, because the characteristic (free-streaming) scale of the standard neutrinos is very large, well into the linear regime (see also~\cite{Fidler:2026wqg} for another recently-proposed general approach).

An important practical obstacle encountered by beyond-CDM attempts, is that evaluating the exact \mbox{1-loop} power spectra for the two coupled fluids is computationally expensive and generally requires numerical integration for all kinematic configurations. However, in~\cite{Verdiani:2025jcf} we showed that in the limits where the nonlinear solutions are analytically tractable, a precise structure of the perturbative series emerges, as a consequence of the Equivalence Principle. This allowed us to devise a prescription to compute the new kernels in an accurate, if approximate, way. The prescription respects the limits discussed above and at the same time permits a computationally efficient evaluation of the loops, suitable for Markov-Chain Monte Carlo (MCMC) inference. 

Concretely, the nonlinear ULA density contrast at order $n$, denoted as $\delta_a^{[n]}$, is evaluated by rescaling the corresponding perturbation for the CDM+baryon fluid by the ratio of their linear transfer functions,
\begin{equation}\label{eq:prescription}
   \delta_a^{[n]}(\mathbf{k}) = \left( \frac{\delta_a^{[1]}(k)}{\delta_{c}^{[1]}(k)} \right) \delta_{c}^{[n]}(\mathbf{k})\,, 
\end{equation}
and analogously for the nonlinear ULA velocity divergence $\theta_a^{[n]}(\mathbf{k})$, which enters the predictions of the galaxy power spectrum multipoles $P_{gg,\hspace{0.2mm}\ell}$ through redshift-space distortions. Operatively, this allows us to compute the 1-loop corrections involving the ULA perturbations using standard momentum integrals with Einstein-de Sitter kernels. 

Beyond the complexity of computing nonlinearities in the presence of scale-dependent growth, since we are dealing with biased tracers we must account for the presence of two distinct underlying matter components, in order to avoid overconfident and potentially biased inference~\cite{Verdiani:2025jcf,Celik:2025wkt}. At the tracer level, we thus adopt a two-fluid bias expansion where the galaxy number density perturbation traces, in general, both the cold and the ULA density fields,
\begin{equation}
   \delta_g\, \supset\, b_c \delta_c + b_a \delta_a  \,.
\end{equation}
One extra linear bias parameter, $b_a$, is therefore present compared to the standard case. To properly capture small deviations from pure $\Lambda$CDM, it is crucial to compute $\delta_a$ up to the appropriate nonlinear order by means of our prescription in Eq.~\eqref{eq:prescription}, avoiding spurious effects that would arise if, e.g.,~it were treated as a strictly linear field. The inclusion of the $\delta_a$ field in the bias expansion demands two new counterterms,
\begin{equation}
P_{gg,\hspace{0.2mm}\ell} (k,z) \supset - 2 c_{ca, \ell} k^2 P_{c a}^L(k,z) - 2 c_{aa, \ell} k^2 P_{a a}^L(k,z)\,,
\end{equation}
in addition to the standard counterterm for the cold component. In principle, the two-fluid treatment also entails several new bias parameters beyond linear level~\cite{Schmidt:2016coo,Chen:2019cfu,Bottaro:2023wkd}; however, these are expected to be poorly constrained by the data, and therefore are not included in our analysis~\cite{Verdiani:2025jcf}. For further details, including the treatment of redshift-space distortions, we refer the reader to~\cite{Verdiani:2025jcf}.

The right panel of Fig.~\ref{fig:ULA-linear-power} shows the measurement of the galaxy power spectrum monopole for the \texttt{LRG1} bin of DESI DR1 (yellow datapoints with errorbars), together with illustrative theoretical predictions for $\maeV{-26}$. The blue line corresponds to the best-fit model discussed in Sect.~\ref{sect:results}, which is compatible with $f_{\rm ax} =0$, while the color gradient shows the effect of increasing the amount of energy density in ULAs. For reference, the dashed cyan line shows the prediction with $f_{\rm ax}$ set to the $95\%$ CL upper bound of the DESI-only analysis. 

\subsection{Parameters and priors}\label{sect:params-priors}

In the analysis we consider a set of ULA mass values spaced logarithmically in the range $[10^{-32},\,10^{-24}]\,\mathrm{eV}$, and for each $m_a$ we vary the parameters reported in Table~\ref{tab:varied-parameters}, see Appendix~\ref{sect:DESI-lcdm}.

Concerning the cosmological parameters, when combining with the CMB we use wide uninformative priors on all of them, as they are well constrained by the data. In the DESI-only analysis, as in~\cite{DESI:2024jxi}, we instead set a Big Bang Nucleosynthesis Gaussian prior on $\Omega_b h^2\sim \mathcal{N}(0.02218,0.00055^2)$, and an informative, but still wide, prior on $n_s\sim \mathcal{N}(0.9649,0.0422^2)$, while fixing $\tau_{\rm reio}=0.0544$. The sum of neutrino masses is always fixed to its minimum allowed value, $\sum m_\nu=0.06\,\mathrm{eV}$. While the possible interplay of light scalar field dynamics and neutrino mass inference is receiving attention in light of current tensions~\cite{Graham:2025dqn}, we note that allowing the neutrino masses to take larger values would yield tighter constraints on the ULA density.

In the baseline analysis, we assume wide Gaussian priors for the bias and counterterm parameters related to the axion component, analogously to the cold fluid. However, when the fraction of non-cold DM is constrained to be very small, one expects that galaxy properties should display a proportionally suppressed sensitivity to this subcomponent~\cite{Verdiani:2025jcf}. Hence, as a proof of principle we also consider $\mathcal{O}(f_\mathrm{ax})$ theoretical priors on the ULA parameters,
\begin{equation}\label{eq:fax-priors}
\begin{split}
    b_a \sim&\, \mathcal{N} \big(0, ( b_c^{\Lambda\mathrm{CDM}}\,f_{\rm ax})^2\big)\,, \\ 
    c_{ca,\ell},\,c_{aa,\ell}\sim&\, \mathcal{N}\big(0, (60 f_{\rm ax})^2 \big)\,,
    \end{split}
\end{equation}
where $f_{\rm ax}$ is fixed at the conservative $95\%$ CL upper bound reported in Table~\ref{tab:result-uppers}, while $b_c^{\Lambda\mathrm{CDM}}$ is the linear bias parameter inferred from a $\Lambda$CDM analysis.
\begin{figure*}[ht]
    \centering
    \begin{subfigure}{.49\textwidth}
     \includegraphics[width=\linewidth]{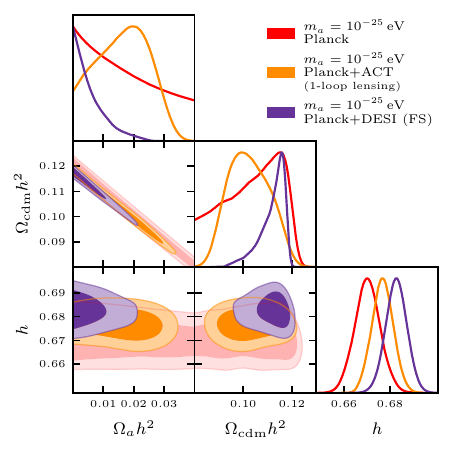}
    \end{subfigure}
    \begin{subfigure}{.49\textwidth}
    \includegraphics[width=\linewidth]{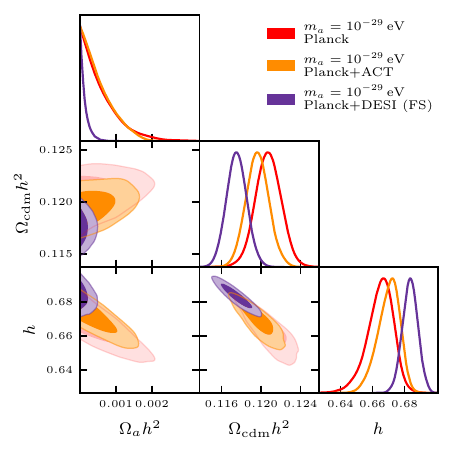}
    \end{subfigure}
    \caption{Marginalized posteriors for Planck alone, Planck + ACT, and Planck + DESI full-shape data. \textit{Left}: $\maeV{-25}$. The ULA field begins to redshift as nonrelativistic matter early in radiation domination, hence $\Omega_a h^2$ shares a strong degeneracy with $\Omega_{\rm cdm} h^2$ and the ULA signatures are mainly in perturbations. To remain agnostic over theoretical systematics, nonlinearities for ACT CMB lensing have been modeled in a perturbative EFTofLSS approach. \textit{Right}: $\maeV{-29}$. The ULA field behaves as a dark energy component until well after recombination, therefore it is mainly constrained by the modification of the distance to last scattering, as evidenced by the degeneracy between $\Omega_a h^2$ and $h$.} 
    \label{fig:m25-m29-corners}
\end{figure*}
\begin{figure*}[ht]
    \centering
    \begin{subfigure}{.325\textwidth}
     \includegraphics[width=2.28in]{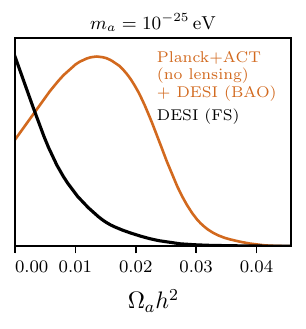}
    \end{subfigure}
    \begin{subfigure}{.325\textwidth}
    \includegraphics[width=2.28in]{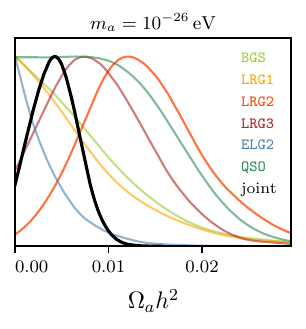}
    \end{subfigure}
    \begin{subfigure}{.325\textwidth}
    \includegraphics[width=2.25in]{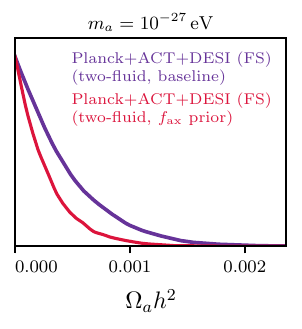}
    \end{subfigure}
    \caption{Results involving the full shape of DESI DR1 galaxy power spectra. \textit{Left}: The full-shape constraint for a model with $\maeV{-25}$ is compared against a combination of ``geometry'' probes given by CMB primaries + BAO. This shows that for the largest ULA masses the shape suppression of the galaxy power spectrum is the prime source of constraining power. \textit{Middle}: Single-redshift-bin results for ULAs with $\maeV{-26}$. Two LRG bins hint at a preference for a non-zero value of $\Omega_a h^2$. However, the statistical significance is mild, gets further reduced in the joint-bin analysis and disappears when combining with the CMB. \textit{Right}: Comparison of the constraints obtained for $\maeV{-27}$ by employing wide agnostic priors, vs.~theoretically-motivated $\mathcal{O}(f_\mathrm{ax})$ priors, on the ULA linear bias parameters and counterterms. See Sect.~\ref{sect:params-priors} for further details.}
    \label{fig:DESIonly-results}
\end{figure*}

\section{Results}\label{sect:results}
\begin{figure*}[t]
    \centering
    \includegraphics[width =0.88\textwidth]{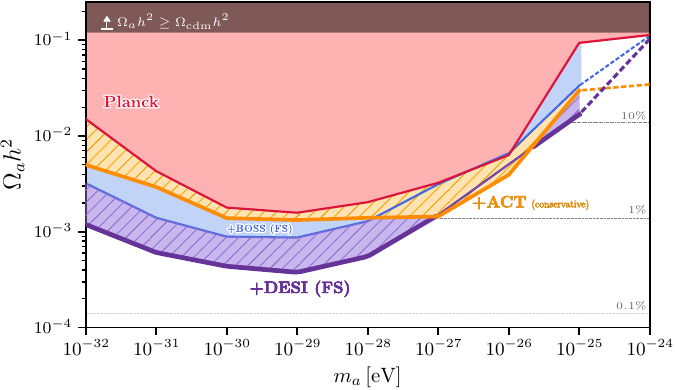}
    \caption{$95\%$ CL upper bounds on the energy density in ULAs today as functions of the ULA mass, obtained in this work from large-scale perturbative probes. The results of DESI, ACT and BOSS are individually combined only with Planck. The hatched areas correspond to the novel exclusions derived in this work. The combined constraints for $m_a > 10^{-25}\,\mathrm{eV}$ are shown by dashed lines, as they are subject to theoretical systematics due to the large allowed value of the ULA fraction $\fax$. The gray horizontal  lines mark the different orders of magnitude for $\fax$.}
    \label{fig:exclusion_plot}
\end{figure*}

The results for the analysis setup discussed above are presented first for the CMB alone, followed by the DESI DR1 full shape as the main novelty of this work, and finally their combination. As we never find a preference for deviations from pure $\Lambda$CDM in the joint analysis, the final results are summarized in Fig.~\ref{fig:exclusion_plot} as an exclusion plot in the $\Omega_a h^2$ vs.~$m_a$ plane. The corresponding numerical values are reported in Table~\ref{tab:result-uppers}.

\paragraph{Planck and ACT DR6} 
Figure~\ref{fig:m25-m29-corners} shows the posterior distributions for the CMB-only dataset given by the combination of Planck and ACT DR6, including lensing. In the left panel we take $\maeV{-25}$, i.e.,~a case where the ULA background evolution transitions to nonrelativistic matter early in radiation domination. As such, lensing is expected to be the main source of constraining power for this data combination (see Fig.~\ref{fig:ULAs_schematic}). As argued in Sect.~\ref{sect:cmb-theory-predictions}, matter nonlinearities are relevant for CMB lensing predictions and when the allowed ULA fraction $\fax$ is rather large, the axion's own nonlinear effects may also become important for properly evaluating the tails of the posterior. For this reason we adopted a treatment based on perturbation theory, which can effectively marginalize over the theoretical systematics due to different halo-model assumptions. As shown in Appendix~\ref{sect:lensing-ULAs-1loop}, we find that the perturbative treatment widens the constraints on $\Omega_a h^2$ (whereas the impact is negligible for the other cosmological parameters). This can be explained by the degeneracy between the ULA energy density and the counterterm: notice that, according to Eq.~\eqref{eq:Pmm-1l-lensing}, a negative $c_m(0)$ counteracts the shape suppression caused by the ULAs. 

We also comment on $\maeV{-24}$, the largest mass considered in our analysis. In this case the ULA subcomponent is even less constrained by our conservative modeling of reconstructed lensing for the \texttt{baseline} scale cut of ACT ($L_{\rm max} = 763$). Nevertheless, we include this point in Fig.~\ref{fig:exclusion_plot} as a dashed line, as an indication of the sensitivity of the dataset, albeit with a possibly large theoretical error. We further note that our Planck+ACT constraints on $\Omega_a h^2$ for $m_a = 10^{-26}\,(10^{-25})\,\mathrm{eV}$ are weaker by a factor of $1.6\,(2.6)$ compared to the results of~\cite{Lague:2026sbd}, which employed the version of~\texttt{axionHMcode} from~\cite{Dome:2024hzq} with an extended scale cut ($L_\mathrm{max}= 1250$). While the analysis presented here is agnostic to different halo-model assumptions, we leave for future work a more direct comparison of the two approaches, for instance by considering an informative prior on the $c_m$ nuisance parameter.

In the right panel of Fig.~\ref{fig:m25-m29-corners} we consider much lighter ULAs, with $\maeV{-29}$, which behave as a dark energy component until $z_{m_a}\approx 250$, well after recombination. In this case, the constraining power is dominated by primary CMB anisotropies, especially through the impact on the angular size of the sound horizon at recombination, $\theta_*$, and on the integrated Sachs-Wolfe effect at very low $\ell$~\cite{Hlozek:2014lca}. The addition of ACT to Planck does not lead to an appreciable change in the axion density marginalized posterior, despite the small effects it entails for other parameters. The visible $\Omega_ah^2\,$-$\,h$ degeneracy direction can be readily interpreted as stemming from keeping $\theta_*$, which is measured very accurately by the CMB primaries, fixed. It is, therefore, a geometric constraint. 

More generally, for $m_a\lesssim 10^{-26}\,\mathrm{eV}$ ACT has additional constraining power compared to Planck-only, as reported by the orange vs.~red exclusions in Fig.~\ref{fig:exclusion_plot}. As noted in~\cite{AtacamaCosmologyTelescope:2025nti}, this is partially driven also by the degeneracy with $n_s$, for which ACT favors a slightly higher value. 

\paragraph{DESI DR1 full shape}

Figure~\ref{fig:DESIonly-results} reports the posterior for $\Omega_a h^2$ obtained from the full shape of DESI DR1 galaxy power spectra, for three masses where DESI is expected to be informative independently of the CMB. Remarkably, as shown in the left panel for $m_a = 10^{-25}\,\mathrm{eV}$, for the heaviest ULAs considered in this work the galaxy power spectrum alone is more constraining than the combination of primary CMB anisotropies and BAO geometric information. This genuinely stems from the fact that the small-scale $P_{gg,\hspace{0.2mm}\ell}$ datapoints probe a suppression of power compared to the amplitude on large scales, and would strongly benefit from reaching even larger $k$, such as could be achieved with the real-space combination $Q_0$~\cite{Ivanov:2021fbu,DAmico:2021ymi}. However, that is beyond the recommended DESI DR1 data cuts and therefore not included here.

For $\maeV{-26}$, the middle panel of Fig.~\ref{fig:DESIonly-results} shows the posterior obtained from single-redshift-bin analyses, as well as their combination. As can be seen, two LRG bins~(\texttt{LRG2} and, albeit less strongly,~\texttt{LRG3}) show a preference for a non-zero value of $\Omega_a h^2$. Interestingly, for the same $m_a$ a similar preference was found in the full-shape analysis of BOSS LRG~\cite{Rogers:2023ezo,Verdiani:2025jcf}. This should not be totally unexpected, as DESI DR1 shares $27\%$ of targets with the BOSS LRG catalog \cite{DESI:2024jxi}. Nevertheless, all other tracers do not exhibit a preference for any deviation from pure CDM and, accordingly, the joint-bin analysis only exhibits a very mild preference, fully compatible with a statistical fluctuation.

\paragraph{Combined}
The combination of CMB and galaxy power spectrum full shape is, of course, expected to be the most informative one. Qualitatively, the heavier ULAs considered in this work are mainly probed via the shape suppression, while CMB primary anisotropies calibrate the standard cosmological parameters. For lighter ULAs, which are already well-constrained by primary CMB anisotropies, galaxy clustering is also able to add further information about the expansion at late times.

We first derive Planck+DESI constraints without including ACT, to highlight the genuine constraining power of the full-shape analysis. These are reported in purple in Fig.~\ref{fig:m25-m29-corners}, showing that both at the high and low ends of the mass range we consider, the improvement from adding DESI is considerable. For $\maeV{-29}$, shown in the right panel of Fig.~\ref{fig:m25-m29-corners}, beyond the non-degenerate information that is added the improvement is also driven by the DESI preference for a higher value of the Hubble parameter. Indeed, along the degeneracy direction at fixed $\theta_*$ imposed by CMB data alone, a higher $h$ is only compatible with a lower $\Omega_a h^2$. 

The right panel of Fig.~\ref{fig:DESIonly-results} shows, for $m_a =  10^{-27}\,\mathrm{eV}$, the further gain in constraining power one can obtain by using the theoretically-motivated priors given in Eq.~\eqref{eq:fax-priors}. This improvement, which is indeed expected to be most relevant in a mass window around $m_a \sim 10^{-27}\,\mathrm{eV}$~\cite{Verdiani:2025jcf}, would be important especially if data were to show a preference for an ULA subcomponent. The validity of such narrower priors may be assessed, for instance, using N-body simulations, as done for massive neutrinos~\cite{Castorina:2015bma, Villaescusa-Navarro:2017mfx, Verdiani:2025znc}. We leave this for future investigation.

\renewcommand{\arraystretch}{1.3}
\begin{table}[b]
    \centering
\begin{tabular}{c|c|c}
%\hline
\multicolumn{3}{c}{\textbf{95\% CL upper limit on} $\,\Omega_a h^2${\tiny{$(f_\mathrm{ax})$}}}\\
%\hline
   $m_a\,[\mathrm{eV}]$&  {Planck + DESI~(FS)} & {Planck + ACT + DESI~(FS)} \\
\hline
$10^{-25}$ & 0.017\tiny{(12\%)} & 0.022\tiny{(16\%)} \\
$10^{-26}$ & 0.0051\tiny{(3.6\%)} &  0.0048\tiny{(3.4\%)} \\
$10^{-27}$ & 0.0015\tiny{(1.1\%)} &  0.0011\tiny{(0.81\%)} \\
$10^{-28}$ & 0.00055\tiny{(0.39\%)} &  0.00064\tiny{(0.46\%)} \\
$10^{-29}$ & 0.00038\tiny{(0.27\%)} &  0.00047\tiny{(0.33\%)} \\
$10^{-30}$ &  0.00044\tiny{(0.31\%)} &  0.00045\tiny{(0.32\%)} \\
$10^{-31}$ &  0.00061\tiny{(0.43\%)} &  0.00063\tiny{(0.45\%)} \\
$10^{-32}$ &  0.0012\tiny{(0.84\%)} &  0.0012\tiny{(0.83\%)} \\
%\hline
\end{tabular}
% $\mathrm{(*)}$ 
\caption{$95\%$ CL upper limits on the fraction of energy density in ULAs, $\Omega_a h^2$. The corresponding upper limit on $f_{\rm ax}$ is expressed as a percentage.}\label{tab:result-uppers}
\end{table}

Finally, Fig.~\ref{fig:exclusion_plot} and Table~\ref{tab:result-uppers} summarize the main results of this work, in the form of upper limits on $\Omega_ah^2$ as functions of the ULA mass. We see that in large portions of the parameter space the addition of DESI data, and of ACT data to a lesser extent, reduce the allowed ULA energy density by more than a factor of two compared to previous analyses. These are among the leading constraints on ULAs in the considered mass range. We do not include here results from probes of smaller scales, whose modeling lies beyond the perturbative regime. These include for example the Lyman-$\alpha$ forest~\cite{Kobayashi:2017jcf, Garcia-Gallego:2025ConstrainingMixed,Rogers:2020ltq,Irsic17,nori19}, which has anyway not been studied in the range of ULA masses relevant to this work, or counts of high-redshift galaxies in the ultraviolet band~\cite{Irsic20,Winch:2024mrt,Sipple:2024svt}. 

\section{Conclusions}
\label{sect:conclusions}

In this paper we presented the first joint analysis of DESI DR1 full-shape and state-of-the-art CMB data aimed at probing a non-cold DM component, taking ULAs as our benchmark scenario.

We performed an independent reanalysis of the DESI DR1 galaxy power spectra, consistently modeling the observables from first principles. Our implementation builds on the two-fluid EFTofLSS framework developed in~\cite{Verdiani:2025jcf}, which allowed us to constrain the signatures of $\Omega_a h^2$ without resorting to phenomenological assumptions (such as on bias parameters) calibrated only on \lcdm\, or massive neutrino cosmologies. Our analysis shows that the combination of galaxy clustering with CMB primary anisotropies and CMB lensing achieves exquisite sensitivity to deviations from pure CDM, restricting the fraction of matter energy density in ULAs, $f_{\rm ax}$, to be below $1\%$ for all masses $m_a \lesssim 10^{-27}\,\mathrm{eV}$, and as small as $0.3\%$ for $m_a \sim 10^{-29}\,\mathrm{eV}$. For the lightest ULAs considered in this work, the addition of DESI DR1 full-shape information to Planck+ACT tightens the bounds on $\Omega_a h^2$ by a factor of $\sim2$, as the galaxy power spectrum breaks the geometric degeneracies that otherwise limit the CMB-only constraints. The DESI DR1 full shape provides new constraining power also at the highest end of the mass range, where it directly probes the suppression of power caused by the ULA Jeans scale: as shown in Fig.~\ref{fig:exclusion_plot} and Table~\ref{tab:result-uppers}, for $m_a = 10^{-25}\,\mathrm{eV}$ the combination of DESI with Planck gives the tightest bound. Our results represent the most stringent limits on the abundance of ULAs from the combination of galaxy clustering and CMB observables, which admit a perturbative modeling.

An intriguing feature of the DESI DR1 full-shape analysis is a mild preference for a non-zero ULA fraction at $m_a\approx10^{-26}\,\mathrm{eV}$, see the middle panel of Fig.~\ref{fig:DESIonly-results}. This is driven primarily by the \texttt{LRG2}, and partially the \texttt{LRG3}, redshift bins and follows a previous hint from BOSS analyses~\cite{Rogers:2023ezo,Verdiani:2025jcf}. However, the preference is substantially diluted  once all DESI tracers are combined, and disappears entirely upon combination with CMB data. We thus interpret it as a statistical fluctuation, rather than a hint of new physics.

Several developments would further sharpen our results and extend their reach. On the galaxy clustering side, at the upper end of the ULA mass window the constraints are driven by the shape suppression and would therefore benefit considerably from the inclusion of information from smaller scales, beyond the $k_\mathrm{max}=0.2\,\hinvMpc$ recommended by the DESI collaboration but still in the perturbative regime. On the CMB side, extending the ACT DR6 lensing analysis to the \texttt{extended} multipole range with $L_{\rm max}=1300$~\cite{ACT:2023ubw,Lague:2026sbd} would add valuable constraining power, particularly for the largest masses, provided the nonlinear modeling of the ULA contribution to the matter power spectrum is brought under sufficient theoretical control. Indeed, pushing this framework to masses above $10^{-25}\,\mathrm{eV}$, where the allowed $f_\mathrm{ax}$ is large enough for genuine ULA nonlinearities to be important, will require a dedicated study. Our 1-loop perturbative treatment of CMB lensing is a first step in this direction, but a more complete analysis will be needed to robustly extend the exclusion limits into that regime. Simulations could also be leveraged, for instance to devise physically motivated priors on the axion bias and counterterm parameters $b_a$, $c_{ca,\ell}$, and $c_{aa,\ell}$, refining the conservative priors adopted here. We leave these directions for future work.

\acknowledgments
FV thanks Benjam\'in Camacho Quevedo, Chiara Moretti, Mohamed Yousry Elkhashab and Pierre Zhang for useful discussions. All the authors thank the Galileo Galilei Institute for Theoretical Physics for hospitality and the INFN for partial support during the completion of this work. FV, EC, ESe and MV are supported by the INFN InDark grant. FV and ESe acknowledge support from the Theory Grant 2023 ``NeuMass'' of the Italian National Institute for Astrophysics (INAF). ES is supported by the grant RYC2023-042775-I, funded by the Spanish Ministry of Science, Innovation and Universities (MCIU) through the Spanish State Research Agency (AEI, 10.13039/501100011033) and by the FSE+.

This research used data obtained with the Dark Energy Spectroscopic Instrument (DESI). DESI construction and operations is managed by the Lawrence Berkeley National Laboratory. This material is based upon work supported by the U.S. Department of Energy, Office of Science, Office of High-Energy Physics, under Contract No.~DE–AC02–05CH11231, and by the National Energy Research Scientific Computing Center, a DOE Office of Science User Facility under the same contract. Additional support for DESI was provided by the U.S. National Science Foundation (NSF), Division of Astronomical Sciences under Contract No.~AST-0950945 to the NSF’s National Optical-Infrared Astronomy Research Laboratory; the Science and Technology Facilities Council of the United Kingdom; the Gordon and Betty Moore Foundation; the Heising-Simons Foundation; the French Alternative Energies and Atomic Energy Commission (CEA); the National Council of Humanities, Science and Technology of Mexico (CONAHCYT); the Ministry of Science and Innovation of Spain (MICINN), and by the DESI Member Institutions: www.desi.lbl.gov/collaborating-institutions. The DESI collaboration is honored to be permitted to conduct scientific research on I’oligam Du’ag (Kitt Peak), a mountain with particular significance to the Tohono O’odham Nation. Any opinions, findings, and conclusions or recommendations expressed in this material are those of the authors and do not necessarily reflect the views of the U.S. National Science Foundation, the U.S. Department of Energy, or any of the listed funding agencies.

\appendix
\section{Details on the perturbation theory modeling of CMB lensing}\label{sect:lensing-ULAs-1loop}

In Sect.~\ref{sect:cmb-theory-predictions} of the main text we have presented our EFTofLSS-based modeling of the CMB lensing power spectrum for $m_a \geq 10^{-26}\,\mathrm{eV}$. In this Appendix, we provide some additional details.  

\begin{figure}[t]
    \centering
    \includegraphics[width=0.9\linewidth]{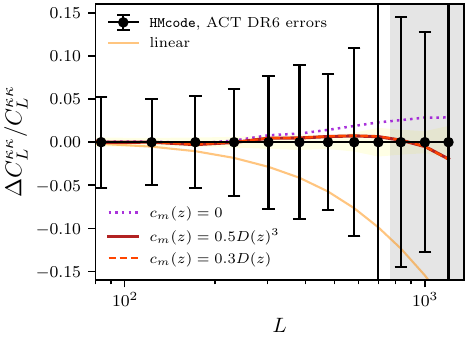}
    \caption{Tests on perturbation theory models for CMB lensing in $\Lambda$CDM, given ACT DR6 errorbars. This assumes Eq.~\eqref{eq:cs2-timedep} for the counterterm, where $c_m(0)$ has been fit to match the \texttt{HMcode} prediction. We display two different values for the exponent, $p = 3$ and $p = 1$, showing that the fit is not very sensitive to this assumption. The linear and the bare 1-loop ($c_m = 0$) predictions are also reported.}
    \label{fig:Ckk-1loop-lcdm}
\end{figure}

For the EFT counterterm $c_m$ we have assumed the time-dependence in Eq.~\eqref{eq:cs2-timedep}, where the value $p=3$ is motivated by numerical studies~\cite{Foreman:2015uva}. We tested that other values of $p$ produce, for what concerns $C^{\kappa\kappa}_L$, an equally good fit to the \texttt{HMcode} prediction when letting $c_m(0)$ free to vary, as shown in Fig.~\ref{fig:Ckk-1loop-lcdm} assuming a  $\Lambda$CDM cosmology. This is expected, as for the range $L < 763$ considered in this work the lensing power spectrum integral is dominated by a limited range in redshift, rendering the precise time dependence of $c_m$ a minor effect. This statement may need to be revised if the lensing analysis is extended to smaller scales, or new data become available with significantly smaller errorbars.

\begin{figure}[b]
    \centering
     \includegraphics[width=\linewidth]{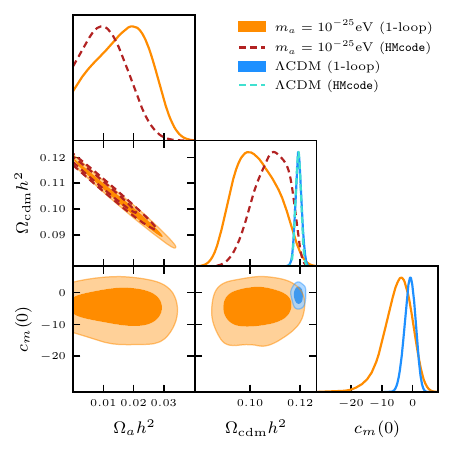}
    \caption{Constraints from Planck+ACT for an ULA cosmology with $\maeV{-25}$ and in $\Lambda$CDM, where nonlinearities for CMB lensing have been modeled with the 1-loop perturbative approach (solid), or computed using the standard~\texttt{HMcode} (dashed). The perturbative modeling requires a counterterm, for which the form in Eq.~\eqref{eq:cs2-timedep} has been assumed, where $p = 3$ while $c_m(0)$ is fit to data.} 
\label{fig:corner_EFTlensing_m25}
\end{figure}

Turning now to the impact on parameter constraints, Fig.~\ref{fig:corner_EFTlensing_m25} reports the results of Planck+ACT analyses in $\Lambda$CDM and $\maeV{-25}$ cosmologies, performed both using~\texttt{HMcode} and applying our EFT modeling for CMB lensing. The $\Lambda$CDM analysis is useful to show that the prior for $c_m(0)$ assumed in this work, see Table~\ref{tab:varied-parameters}, is wide enough and the posterior is constrained by the data. The $\maeV{-25}$ results show that, as expected, the 1-loop analysis yields broader constraints compared to computing nonlinearities with~\texttt{HMcode}, which does not have nuisance parameters. As argued in Sect.~\ref{sect:results}, our 1-loop analysis effectively marginalizes over different theoretical assumptions on ULA nonlinearities, keeping this systematic uncertainty under control.

\section{Details on the DESI DR1 full-shape analysis}\label{sect:DESI-lcdm}
In the DESI full-shape analysis of galaxy clustering, we fit the monopole and quadrupole of the galaxy power spectrum up to $k_\mathrm{max}=0.2\,\hinvMpc$ for all six redshift bins. This matches the baseline choice of previous analyses \cite{DESI:2024jxi,DESI:2024hhd} and \cite{Chudaykin:2025aux}; the latter reference also includes the hexadecapole, which however has a very low signal-to-noise ratio. We do not perform a combination with post-reconstruction BAO information.

\begin{table*}[!t]
    \centering  
    \begin{tabular*}{.8\textwidth}{@{\extracolsep{\fill}} c c c c ccc }
    %\hline
        \multicolumn{7}{c}{\textbf{Cosmological}}\\
        \hline
        $\Omega_a h^2$ & $\Omega_{\rm cdm} h^2$ & $\Omega_b h^2$ &  $h$ & $\log(10^{10}A_s)$ & $n_s$ & $\tau_{\rm reio}$ \\
       % \hline
    \end{tabular*}

    \vspace{1em} % Vertical gap

    \begin{minipage}[t]{0.39\textwidth}
    \hfill
        \begin{tabular*}{\textwidth}{@{\extracolsep{\fill}} cc }
            %\hline
            \multicolumn{2}{c}{\textbf{CMB nuisance} (sampled)}\\[-.3em]
            \scriptsize{parameter} & \scriptsize{prior/value}\\
            \hline
            $A_\mathrm{ACT}$& $\mathcal{N}(1,0.03^2)$\\
            $p_\mathrm{ACT}$ & $[0.9,1.1]$\\
            $A_\mathrm{Planck}$& $A_\mathrm{ACT}$\\
            $c_m(0)$& $\mathcal{N}(0,10^2)$\\
            %\hline
        \end{tabular*}
    \end{minipage}
    \hspace{.01\textwidth}
    \begin{minipage}[t]{.39\textwidth}
        \begin{tabular*}{\textwidth}{@{\extracolsep{\fill}} cc }
            %\hline
            \multicolumn{2}{c}{\textbf{EFTofLSS nuisance} (sampled)}\\[-.3em]
            \scriptsize{parameter} & \scriptsize{prior}\\
            \hline
            $b_c\sigma_{8,c}$& $[0,3]$\\
            $b_a\sigma_{8,c}$& $[-3,3]$\\
            $b_2\sigma_{8,c}^2$&$\mathcal{N}(0,5^2)$\\
            $b_{\mathcal{G}_2}\sigma_{8,c}^2$ &$\mathcal{N}(0,5^2)$\\
            %\hline
        \end{tabular*}
    \end{minipage}

    \vspace{1em} % Vertical gap

    \begin{tabular*}{.8\textwidth}{@{\extracolsep{\fill}} cccc }
    %\hline
        \multicolumn{4}{c}{\textbf{EFTofLSS nuisance} (analytically marginalized)}\\
        %\hline
        \scriptsize{parameter} & \scriptsize{prior} & \scriptsize{parameter} & \scriptsize{prior} \\
        \hline
        $c_{cc,\ell}$ & $\mathcal{N}(0,30^2)_{\ell\hspace{0.1mm} = \hspace{0.1mm}0}\,,\; \mathcal{N}(30,30^2)_{\ell \hspace{0.1mm}=\hspace{0.1mm}2}$  & $b_{\Gamma_3}$& $\mathcal{N}(0,5^2)$ \\
        $c_{ca,\ell}$ & $\mathcal{N}(0,60^2)$ & $P_{\rm shot}$ & $\mathcal{N}(0,1^2)$ \\
        $c_{aa,\ell}$ & $\mathcal{N}(0,60^2)$ & $a_2$ & $\mathcal{N}(0,5^2)$ \\
        %\hline
    \end{tabular*}

    \caption{Parameters and priors used in the analysis, for each assumed value of the ULA mass $m_a$. For the cosmological parameters we always use wide uninformative priors, with the exception of $\Omega_b h^2$, $n_s$ and $\tau_{\rm reio}$ when not combining with the CMB,  as in \cite{DESI:2024jxi}. The counterterms $c_m(0), c_{cc,\ell}, c_{ca,\ell}$ and $c_{aa,\ell}$ have dimensions of $h^{-2}\, \mathrm{Mpc}^2$. Since we fit the monopole and quadrupole of the galaxy power spectrum, the index $\ell$ in the table runs over $0$ and $2$.}\label{tab:varied-parameters}
\end{table*}

Our theoretical model is based on the EFTofLSS~\cite{Baumann:2010tm,Carrasco:2012cv,Lewandowski:2015ziq} and is given, as a function of the wavenumber $k$, by
\begin{equation}
    P_{gg,\hspace{0.2mm}\ell} = P_{gg,\hspace{0.2mm}\ell}^L+ P_{gg,\hspace{0.2mm}\ell}^\mathrm{1\text{-}loop}+P_{gg,\hspace{0.2mm}\ell}^\mathrm{ctr}+P_{gg,\hspace{0.2mm}\ell}^\mathrm{SN}\,,
\end{equation}
where the 1-loop contribution is computed with the same bias expansion and counterterms as in~\cite{Verdiani:2025jcf}. Concerning the stochastic contribution, before expanding in multipoles it is given by~\cite{DESI:2024jxi} 
\begin{equation}
  P_{gg}^\mathrm{SN}(k,\mu) = \frac{1}{\bar n}\left[P_{\rm shot} +a_2\hspace{0.3mm} \mu^2 \left(\frac{k}{k_\mathrm{NL}}\right)^{\hspace{-1mm}2}\, \right]\,,
\end{equation}
as the Poissonian prediction is already subtracted from the measurements. Here $\mu \equiv {\bf k}\cdot \widehat{\bf z}/k$, where $\widehat{\bf z}$ is the unit vector along the line of sight. We set $k_\mathrm{NL}=0.45\,\hinvMpc$, although its exact value is not important, as $a_2$ is marginalized over with a wide prior.

The full list of parameters is reported in Table~\ref{tab:varied-parameters}, along with the priors employed. They are divided in three classes: cosmological, sampled nuisance, and analytically-marginalized nuisance parameters. The theoretical model is then convolved with the survey window function, accounting also for the $\theta$-cut, and compared to data with a Gaussian likelihood. We employ the covariance numerically computed from the \texttt{EZmocks} mocks \cite{Chuang:2014vfa} and rescaled by the appropriate factor to match the analytic-based estimates \cite{DESI:2024jxi, DESI:2024aax}. The likelihood is computed within the \texttt{PBJ}\footnote{https://chiaramoretti.gitlab.io/pbj}~\cite{Moretti:2023drg} code, embedded in the \texttt{Cobaya} framework \cite{Torrado:2020dgo} for the purpose of combination with the other datasets, and then sampled with the \texttt{Nautilus} sampler \cite{Lange:2023ydq}.

We tested our pipeline by comparing {\lcdm} results against both the official~\cite{DESI:2024jxi} and independent~\cite{Chudaykin:2025aux} analyses, finding remarkable consistency, both for single redshift bins as well as for the joint case.

\bibliography{references}

\end{document}